\documentclass[%
 reprint,
superscriptaddress,
preprintnumbers,
nofootinbib,
 amsmath,amssymb,
 aps,
]{revtex4-2}

\usepackage{graphicx}
\usepackage{dcolumn}
\usepackage{bm}
\usepackage{hyperref}
\usepackage{aas_macros}
\usepackage{soul}

\usepackage[usenames,dvipsnames]{xcolor}
\usepackage{hyperref}
\hypersetup{
    colorlinks = true,
    citecolor = {magenta},
    linkcolor = {blue},
    urlcolor = {blue}		
}
\usepackage{mathtools}
\usepackage{subcaption}
\usepackage{transparent}

\usepackage{fontawesome5}
\usepackage{hyperref}

\makeatletter
\newcommand{\github}[1]{%
   \href{#1}{\faGithubSquare}%
}
\makeatother

\def\bi#1{\hbox{\boldmath{$#1$}}}

\begin{document}

\title{
Super-sample covariance of the power spectrum,  \\ bispectrum, halos, voids, and their cross covariances
}

\author{Adrian E.~Bayer}
\email{abayer@berkeley.edu}
\affiliation{
 Berkeley Center for Cosmological Physics, University of California,
Berkeley,  CA 94720, USA
}%
\affiliation{Department of Physics, University of California,
Berkeley,  CA 94720, USA
}%
\affiliation{Lawrence Berkeley National Laboratory,  1 Cyclotron Road, Berkeley, CA 94720, USA}
\affiliation{
Center for Data-Driven Discovery, Kavli IPMU (WPI), UTIAS, The University of Tokyo, Kashiwa, Chiba 277-8583, Japan
}

\author{Jia Liu}
\email{jia.liu@ipmu.jp}
\affiliation{
Center for Data-Driven Discovery, Kavli IPMU (WPI), UTIAS, The University of Tokyo, Kashiwa, Chiba 277-8583, Japan
}

\author{Ryo Terasawa}
\affiliation{
Center for Data-Driven Discovery, Kavli IPMU (WPI), UTIAS, The University of Tokyo, Kashiwa, Chiba 277-8583, Japan
}
\affiliation{
Department of Physics, School of Science, The University of Tokyo, 7-3-1 Hongo, Bunkyo-ku, Tokyo 113-0033, Japan
}

\author{Alexandre Barreira}
\affiliation{
Excellence Cluster ORIGINS, Boltzmannstraße 2, 85748 Garching, Germany
}
\affiliation{
Ludwig-Maximilians-Universität, Schellingstraße 4, 80799 München, Germany
}

\author{Yici Zhong}
\affiliation{
Department of Physics, School of Science, The University of Tokyo, 7-3-1 Hongo, Bunkyo-ku, Tokyo 113-0033, Japan
}
\affiliation{Department of Astrophysical Sciences, Peyton Hall, Princeton University, Princeton, NJ 08544, USA}

\author{Yu Feng}
\affiliation{
Berkeley Center for Cosmological Physics, University of California,
Berkeley,  CA 94720, USA
}%

\date{\today}

\begin{abstract}
We study the effect of super-sample covariance (SSC) on the power spectrum and higher-order statistics: bispectrum, halo mass function, and void size function. We also investigate the effect of SSC on the cross covariance between the statistics. We consider both the matter and halo fields. Higher-order statistics of the large-scale structure contain additional cosmological information beyond the power spectrum and are a powerful tool to constrain cosmology. They are a promising probe for ongoing and upcoming high precision cosmological surveys such as DESI, PFS, Rubin Observatory LSST, Euclid, SPHEREx, SKA, and Roman Space Telescope. Cosmological simulations used in modeling and validating these statistics often have sizes that are much smaller than the observed Universe. Density fluctuations on scales larger than the simulation box, known as super-sample modes, are not captured by the simulations and in turn can lead to inaccuracies in the covariance matrix.  We compare the covariance measured using simulation boxes containing super-sample modes to those without. We also compare with the Separate Universe approach. We find that while the power spectrum, bispectrum and halo mass function show significant scale- or mass-dependent SSC, the void size function shows relatively small SSC. We also find significant SSC contributions to the cross covariances between the different statistics, implying that future joint-analyses will need to carefully take into consideration the effect of SSC. 
To enable further study of SSC, our simulations have been made publicly available at \href{https://github.com/HalfDomeSims/ssc}{\faGithub}.

\end{abstract}

\maketitle


\section{Introduction}
\label{sec:intro}

Ongoing and upcoming cosmological missions such as
DESI\footnote{\url{https://www.desi.lbl.gov}} \citep{collaboration2016desi}, PFS\footnote{\url{https://pfs.ipmu.jp/index.html}} \citep{Takada_2014},
Rubin Observatory LSST\footnote{\url{https://www.lsst.org}} \cite{LSSTSci},
Euclid\footnote{\url{https://www.euclid-ec.org}} \cite{Euclid}, 
SPHEREx\footnote{\url{https://www.jpl.nasa.gov/missions/spherex}} \cite{SphereX_2014}, 
SKA\footnote{\url{https://www.skatelescope.org}} \cite{SKA_2009}, and Roman Space Telescope\footnote{\url{https://roman.gsfc.nasa.gov}} \cite{spergel2013widefield} will probe ever larger volumes of cosmic structure in the small-scale, nonlinear regime.  These data contain rich information that can be used to constrain fundamental physics, such as dark energy, dark matter, and neutrino mass. 
To fully realize the potential of these surveys, many higher-order (or non-Gaussian) statistics have been proposed to extract additional information beyond the power spectrum (2-point function). These include, for example, the bispectrum (3-point function), halo mass function, void size function, probability distribution function, marked power spectrum, and wavelet scattering transform~\cite{Takada_2004, Sefusatti_2006, Berge_2010, Kayo_2013, Schaan_2014, Liu2015,Liux2015,Kacprzak2016,Shan2018,Martinet2018,liu&madhavacheril2019, Li2019, Kreisch2019, Coulton2019, Sahl_n_2019, Marques2019,ajani2020, Hahn_2020, hahn2020constraining, Dai_2020, Uhlemann_2020, Allys_2020, Gualdi_2020,Harnois-Deraps2020, Arka_2020, Massara_2020, Cheng_2020, Cheng_2021, bayer2021detecting, Kreisch_2021,Bayer_2022_fake, Valogiannis_2021, Valogiannis_2022, Eickenberg_2022}. They have been studied intensively in recent years and are becoming standard tools for cosmological inferences. Moreover, there is increased interest in joint analysis, in which second and higher-order statistics are combined to maximize the information gain (see e.g.~\cite{bayer2021detecting, Hamaus_2020, Dvornik_2022, Pandey_2022, DES_2022_LCDM+}).

Models for higher-order statistics usually rely on simulations for validation of analytic theories, calibration of semi-analytic models, or as the base of simulation-based inferences. To compute the covariance matrix of higher-order statistics, one typically requires a large set of simulations with different random initial conditions \cite{quijote}. Such simulations assume periodic boundary conditions
and are normally much smaller than the typical observed volumes of the Universe.
Importantly, the mean density of these simulations is the cosmic one, and so by construction, they do not take into account the effects of perturbations with wavelengths longer than the size of the simulation. These so-called ``super-sample modes'' can however contribute sizeably to the covariance matrix; this effect is called the super-sample covariance (SSC) effect and must be carefully included to achieve accurate results.

To make contact with past literature, SSC has been studied as the ``beat-coupling'' (BC) effect in the mildly nonlinear regime using perturbation theory, as ``halo sample variance'' (HSV) in the highly nonlinear regime using the halo model, and was sometimes called the DC mode effect as an analogy between the constant background fluctuation and constant electric Direct Current~\cite{Sirko_2005, Gnedin_2011}. 
It was first studied in the context of the power spectrum~\cite{Hamilton_2006, Takada_2013}, and its effects have since been quantified using direct simulations~\cite{dePutter_2012}, perturbation theory~\cite{Takahashi_2009, Baldauf_2011, Barreira_2017_0, Barreira_2017}, and separate universe simulations~\cite{Li_2014, Wagner_2015, Terasawa_2022}.
It has also been studied in relation to cluster counts~\cite{Hu_2003, Schaan_2014, Philcox_2020_ehm}, the matter bispectrum~\cite{Sefusatti_2006, Chan_2018, Barreira_2019_squeze, Coulton_2022}, the matter one-point probability density function~\cite{Uhlemann_2022}, the redshift space galaxy power spectrum~\cite{Akitsu_2017,Akitsu_2018,Li_2018}, and the lensing power spectrum~\cite{Barreira_2018}. Furthermore, the effects of baryons on the SSC has been studied in~\cite{Mohammed_2014, Barreira_2019, Anik_2022}. 
The SSC effect associated with gravitational potential perturbations in cosmological with local primordial non-Gaussianity (i.e.~$f_{\rm NL} \neq 0$) has also recently been studied~\cite{Castorina_2020}.
Fast, approximate, methods exist to account for the SSC in forecasts for upcoming lensing and photometric surveys  \cite{Lacasa_2019, Beauchamps_2022, Lacasa_2022}.



In this paper, we study the effects of SSC for the power spectrum and several higher-order statistics: the bispectrum, halo mass function, and void size function. To do so, we compare the statistics measured using small periodic boxes, which ignore SSC, to those using equally-sized boxes that are embedded in a much larger simulation, which include SSC. 
We study the effect of SSC in both the total matter 
field and the halo field. We also validate our results against the separate universe (SU) approach, in which the SSC contribution is calculated semi-analytically using the response of the statistics to certain changes in the cosmological parameters.

Our work is the first to investigate the effects of SSC for voids. Cosmic voids have been studied intensively in recent years~\cite{Platen_2008, Bos_2012, Sutter_2012, Jennings_2013, Pisani_2015, Paillas2017, Sahl_n_2019, Contarini_2019, Verza_2019, Kreisch2019, bayer2021detecting, Kreisch_2021} and have achieved cosmological constraints with observational data~\cite{Hamaus_2020, Hamaus_2022, Contarini_2022, Bonici_2022}. The bias parameters of voids have also been recently studied using SU simulations~\cite{Jamieson_2019, Chan_2020}. Our work is also the first to study the effects of SSC on the cross covariance between the combinations of all of these statistics.

The paper is organized as follows. Section \ref{sec:method} outlines the methods employed to compute the SSC of the power spectrum, bispectrum, halo mass function, and void size function. Section \ref{sec:results} presents the results for SSC of these statistics and their cross covariances. We conclude in Section \ref{sec:conclusions}.

\section{Method}
\label{sec:method}

In this section, we describe the methods used to run the N-body simulations, to compute the statistics, and to compute the SSC. We also briefly describe the SU approach.

\subsection{Covariance}
\label{sec:method_cov}

The covariance matrix between an observable $\mathcal{O}_\alpha$ and another observable $\mathcal{O}_\beta$ is given by
\begin{equation}
C_{\alpha\beta}=\left\langle\left( \mathcal{O}_\alpha - \langle \mathcal{O}_\alpha \rangle \right)\left( \mathcal{O}_\beta - \langle \mathcal{O}_\beta \rangle \right)\right\rangle,
\label{eqn:covm}
\end{equation}
where $\langle\rangle$ denotes the mean over realizations. 
The $\alpha$ and $\beta$ subscripts can refer to different bins of particular statistic, or two completely different statistics.
The covariance can be estimated by evaluating Eqn.~\ref{eqn:covm} using an ensemble of simulations with different random realizations of the initial conditions.

We quantify the SSC effect by comparing the following two sets of simulations:
\begin{itemize}
    \item \textbf{sub}-boxes that are embedded in a much larger simulation, where the effect of SSC is properly captured;
    \item\textbf{small} boxes that are of the same size and resolution as the sub-boxes, but are independently simulated with periodic boundary conditions and have no super-sample modes.
\end{itemize}

Because the super-sample modes are only present in the former  and not in the latter, the SSC is given by
\begin{equation}
    C_{\rm SSC} = C_{\rm sub}-C_{\rm small},
    \label{eq:C_SSC}
\end{equation}
where $C_{\rm sub}$ is the covariance computed using sub-boxes and $C_{\rm small}$ is that using small boxes.

\subsection{N-body simulations}
\label{sec:method_sims}

We use \texttt{FastPM}~\cite{Feng2016, Bayer_2021_fastpm}, a particle-mesh~(PM) N-body simulation, to simulate a big box of side-length $5\, {\rm Gpc} /h$ with $2048^3$ matter particles. We then split this big box into $8^3 = 512$ sub-boxes. We compare these to $512$ independent, periodic, small boxes of size $5000 / 8$=$625 \,{\rm Mpc} / h$, each with $256^3$ particles. The resolution of the small boxes is chosen to match the big box. 
We consider a maximum scale cut of $k_{\rm max}=0.8\,h/{\rm Mpc}$ as scales with lower $k$ than this are well modeled by our simulations. 
In all cases, we begin the simulations at $z=9$ and take 60 steps to $z=0$. The resolution of the force mesh is 2 times the number of particles. For simplicity, we consider and discuss only the results at $z=0$, which is when SSC is expected to be the strongest.

Our cosmological parameters are 
$h=0.6774$, 
$\Omega_m = 0.3089$, 
$\Omega_b = 0.0486$,
$\sigma_8 = 0.8159$,
$n_s = 0.9667$,
$M_\nu=0$.


To identify halos, we use the Friends-of-Friends (FoF) algorithm with a linking length of 0.2. 
We generate the particle and halo overdensity fields using the Cloud-in-Cell (CIC) method with $N_{\rm mesh} = 256$ using \texttt{nbodykit}~\cite{Hand_2018}. Further, we consider the halo field in real and not in redshift space. 
We compensate the field for window effects before calculating the statistics~\cite{Jing_2005}. 
For matter the overdensity field is computed as $\delta=(\rho-\bar{\rho})/\bar{\rho}$, where $\rho$ is the matter density, while for halos it is computed as $(n-\bar{n})/\bar{n}$, where $n$ is the number density.
In order to compute the overdensity field in the sub-boxes, there are two choices for the mean density $\bar{\rho}$ (or $\bar{n}$): using the ``{\bf global}'' mean of the big box, or the ``{\bf local}'' mean of the sub-box.
Realistically, for weak lensing surveys it is appropriate to use the global mean as the mean density can be directly calculated from the cosmological model, moreover, the measured weak-lensing shear field is sensitive to the global-mean density. However, for galaxy surveys, since one does not know how to predict from first principles the total number of galaxies, the local mean is what is most appropriate as we measure the galaxy statistics w.r.t.~the observed galaxy number density in the survey. We consider both cases in our analysis. 

\subsection{Statistics}

\begin{figure*}[t]
\centering
  \begin{subfigure}[t]{\textwidth}
    \includegraphics[width=1\textwidth]{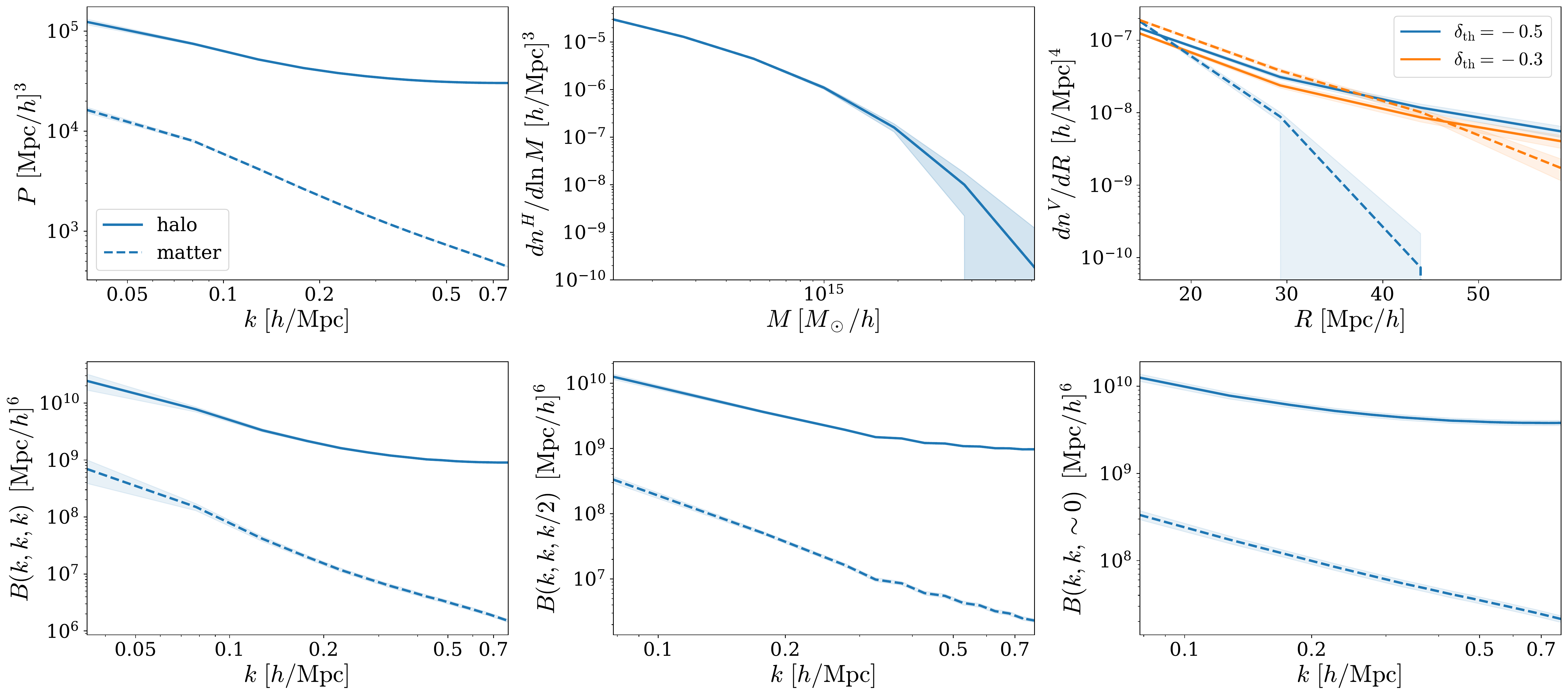}
  \end{subfigure}
  \caption{Power spectrum (top left), halo mass function (top middle), void size function (top right), and three bispectrum configurations (bottom) averaged over all small boxes. Error bands correspond to the standard deviation across the small boxes, $\sqrt{C_{\rm small}}$. We show results for the halo (solid) and matter (dashed) field. For the void size function, we show results for two density thresholds used for void searching, $\delta_{\rm th}=-0.5$ (blue) and $\delta_{\rm th}=-0.3$ (orange).}
\label{fig:stats} 
\end{figure*}

We now define the various statistics considered in this work, and the method used to compute them. 
We plot the statistics averaged over all the sub-boxes in Fig.~\ref{fig:stats}.

\begin{itemize}
    \item The \textbf{power spectrum} $P(k)$ is defined as the Fourier transform of the 2-point correlation function $\xi(x_1-x_2)\equiv\langle \delta(x_1)\delta(x_2) \rangle$, where $\delta$ is the overdensity field. Defining the fundamental frequency of the box as $k_F \equiv 2\pi/625\,h/{\rm Mpc}$, we use 15 linear bins between bin edges [0, 80$k_F$], with $\Delta k = 5 k_F$. 
    
    \item The \textbf{bispectrum} $B(k_1,k_2,k_3)$ is defined as the Fourier transform of the 3-point correlation function  $\langle \delta(x_1)\delta(x_2)\delta(x_3)\rangle$. We consider three particular configurations of the bispectrum: equilateral ($k_1=k_2=k_3$), isosceles ($k_1=k_2\neq k_3$), and squeezed ($k_1=k_2, k_3\sim0$). More concretely, we consider the squeezed mode as $k_3=3.6 \times 10^{-2} \,h/{\rm Mpc}$.
    We compute the bispectrum using the \texttt{bskit} package~\cite{Foreman_2020}, which employs the FFT-based bispectrum estimators of~\cite{Scoccimarro_2000, Sefusatti_2016}. The $k$ binning is the same as for the power spectrum. 
    \item The \textbf{halo mass function} (HMF), denoted $dn^H/d \ln M$, 
    is defined as the comoving number density of halos $n^H$ per unit of log halo mass $\ln M$.
    We consider 7 logarithmic bins bounded by $M_{\rm min} = 10^{14} M_\odot/h$ and $M_{\rm max} = 10^{16} M_\odot/h$.
    
    \item The \textbf{void size function} (VSF), denoted $dn^V/d R$, 
    corresponds to the comoving number density of voids $n^V$ per unit of void radius $R$. 
    We consider spherical voids in smoothed density fields. The $N_{\rm mesh}=256^3$ field is smoothed with top-hat filters of size $R_{\rm filter}$ in 7 linear bins between $R_{\min}=12.2 \,{\rm Mpc}/h$ and $R_{\rm max}=100.1\,{\rm Mpc}/h$, with $\Delta R=6 d_{\rm grid}$ where $d_{\rm grid}\approx2.44 \,{\rm Mpc}/h$ is the grid size.
    We search hierarchically, first finding the largest voids and then the, more abundant, smaller voids.
    Voids of size $R=R_{\rm filter}$ are defined as local minima in the $R_{\rm filter}$-filtered field, with values lower than a predefined threshold $\delta_{\rm th}$, unless they overlap with existing larger voids.
    In this work we investigate thresholds of $\delta_{\rm th}=-0.3$ and $-0.5$.
    The void finding algorithm was developed by \cite{Arka_2016} and we use the implementation in Pylians3 
    \cite{Pylians_2018}.
\end{itemize}

\subsection{Separate universe simulations}
\label{subsec:su}
We now briefly summarize an alternative method to compute the SSC using Separate Universe (SU) simulations; we refer the reader to \cite{Li_2014, Wagner_2015} for more details. In this approach, the effect of a super-survey density mode that is constant inside the box and has amplitude $\delta_b$ is mimicked by adjusting the cosmological parameters such that $\bar{\rho}_m \to \bar{\rho}_m\left(1 + \delta_b\right)$, where $\bar{\rho}_m$ is the mean physical matter density; if the fiducial cosmology is a spatially flat universe, this implies that the separate universe has non-zero curvature ($\Omega_k \neq 0$). The response of any summary statistic $\mathcal{O}$ to $\delta_b$ is computed by considering the difference between simulations run with different $\delta_b$. The SSC is then approximated by
\begin{align}
    C_{\rm SSC-SU}^{ij} \simeq \sigma_b^2 \frac{d \mathcal{O}_i}{d \delta_b} \frac{d \mathcal{O}_j}{d \delta_b},
    \label{eqn:C_SU}
\end{align}
where $\sigma_b^2$ is the variance of the linear matter density fluctuations on the size of the survey described by a window function $W$,
\begin{equation}
    \sigma_b^2 \equiv \frac{1}{V_W^2}\int \frac{d^3 \bi{k}}{(2\pi)^3} | W(\bi{k})|^2 P_{\rm lin}(k),
    \label{eqn:sigma_b}
\end{equation}
where $V_W=\int d^3 \bi{x}~ W(\bi{x})$ is the survey volume and $P_{\rm lin}(k)$ is the linear matter power spectrum.
The window function used in this work corresponds to a 3d cube of side-length $625 \,{\rm Mpc}/h$, giving $\sigma_b^2=6.8\times10^{-5}$.
Concretely, we evaluate the responses using finite difference methods on simulations with $\delta_b = \pm 0.03$, and averaging over 20 realizations of the initial conditions.

The $\delta_b$ mode modifies the background expansion history, which implies some care when choosing the box size of the simulations with $\delta_b \neq 0$. In our simulations here, we choose the comoving box size to match at all times in ${\rm Mpc}$ units. This corresponds to the ``growth-dilation'' methods in the notation of \cite{Li_2014}, or equivalently, with our SU simulations we measure the so-called ``growth-only'' responses in the language of \cite{Wagner_2015}.
Importantly, when identifying halos in the simulations, the FoF linking length in the separate universe needs to be rescaled by the ratio of the scale factors in the two simulations to guarantee matching halo definitions (see e.g.~\cite{Barreira:2020kvh} for a discussion).

Since the simulations with different $\delta_b$ values can be initialized with the same random phases of the initial conditions, the SU approach has the significant advantage of converging with much fewer simulations than the sub-box approach (discussed in Section \ref{sec:method_sims}). We note, however, that our SU simulations account only for the impact of isotropic density perturbations as super-survey modes, i.e., they do not account in particular for the effect of super-survey tidal fields \cite{Schmidt_2018, Masaki_2020, Akitsu_2021}. Here, we consider angular averaged spectra in real space, for which the impact of super-survey tidal fields averages out, but we note that for analyses in redshift space \cite{Akitsu_2017, Li_2018, Akitsu_2018} and weak-lensing applications \cite{Barreira_2018} this is not the case and the super-survey tidal fields can have a non-negligble effect. Further, the super-survey tidal fields contribute also to the SSC effect of halo and void counts, although in a weak manner since this happens only at second order -- this is because the tidal field is a tensor, and thus must be contracted at least once (forming a quadratic quantity) if it is to affect any scalar statistic like counts \cite{Saito:2014qha}. On the other hand, the SSC calculated using the sub-box approach automatically includes both the effects of density and tidal fields.


\section{Results}
\label{sec:results}

Here we show the effect of SSC for individual statistics as well as their cross covariances. In all plots, error bars are computed using bootstrapping and correspond to the $95\%$ confidence interval.

\subsection{Matter field statistics}
\label{subsec:cdm}

\begin{figure*}[t]
\centering
  \begin{subfigure}[t]{\textwidth}
    \includegraphics[width=0.32\textwidth]{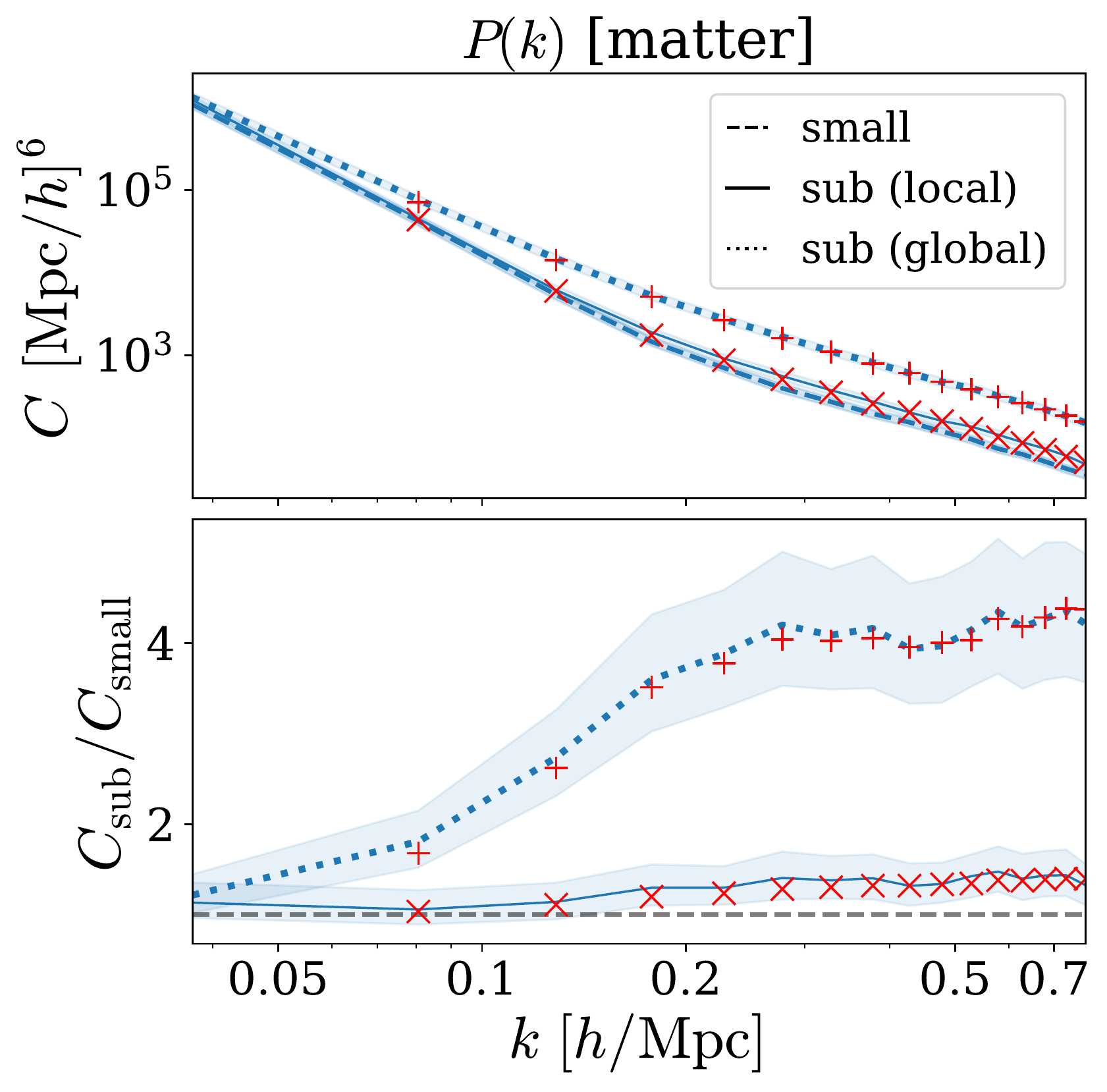}
    \includegraphics[width=0.32\textwidth]{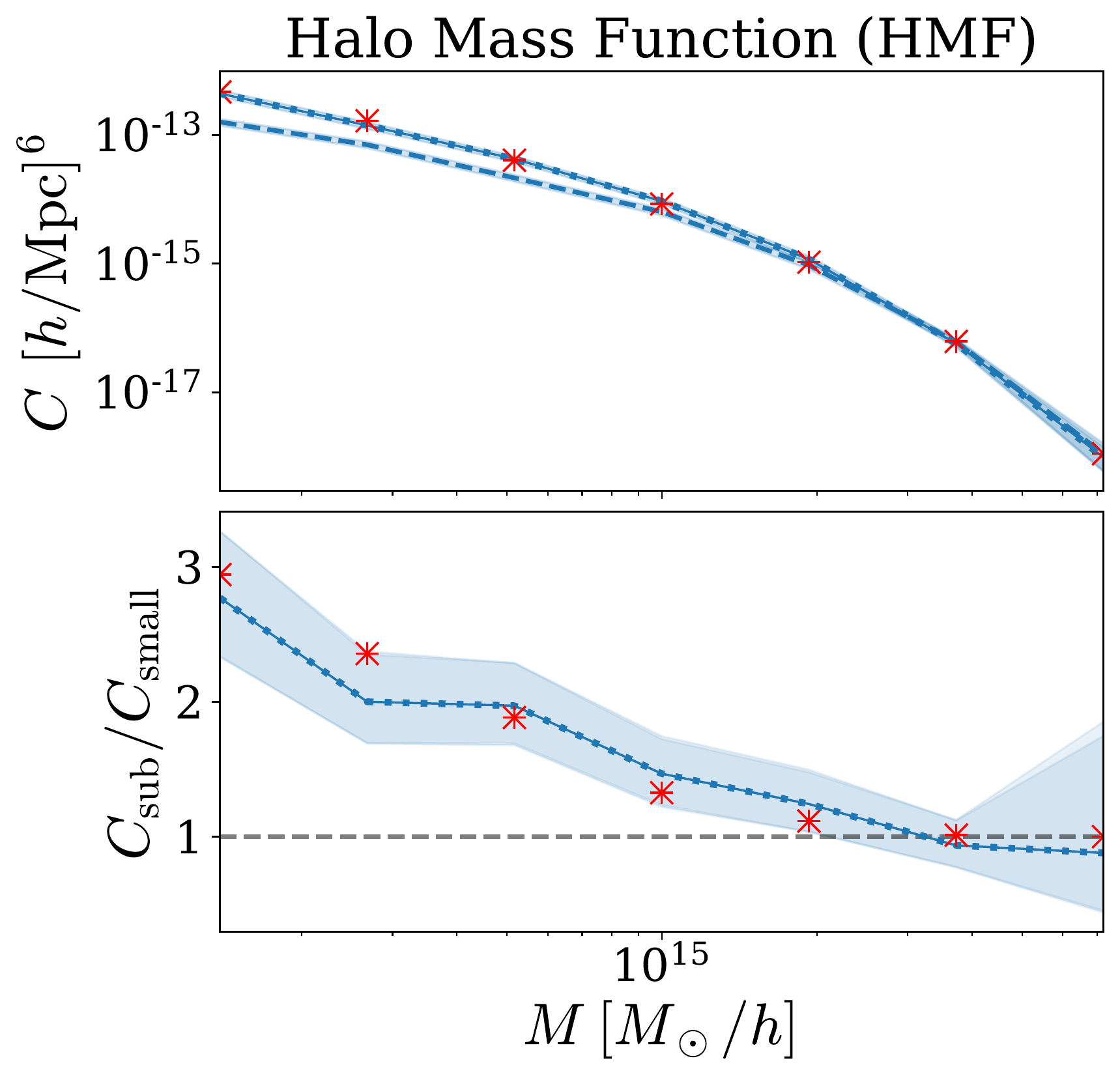}
    \includegraphics[width=0.32\textwidth]{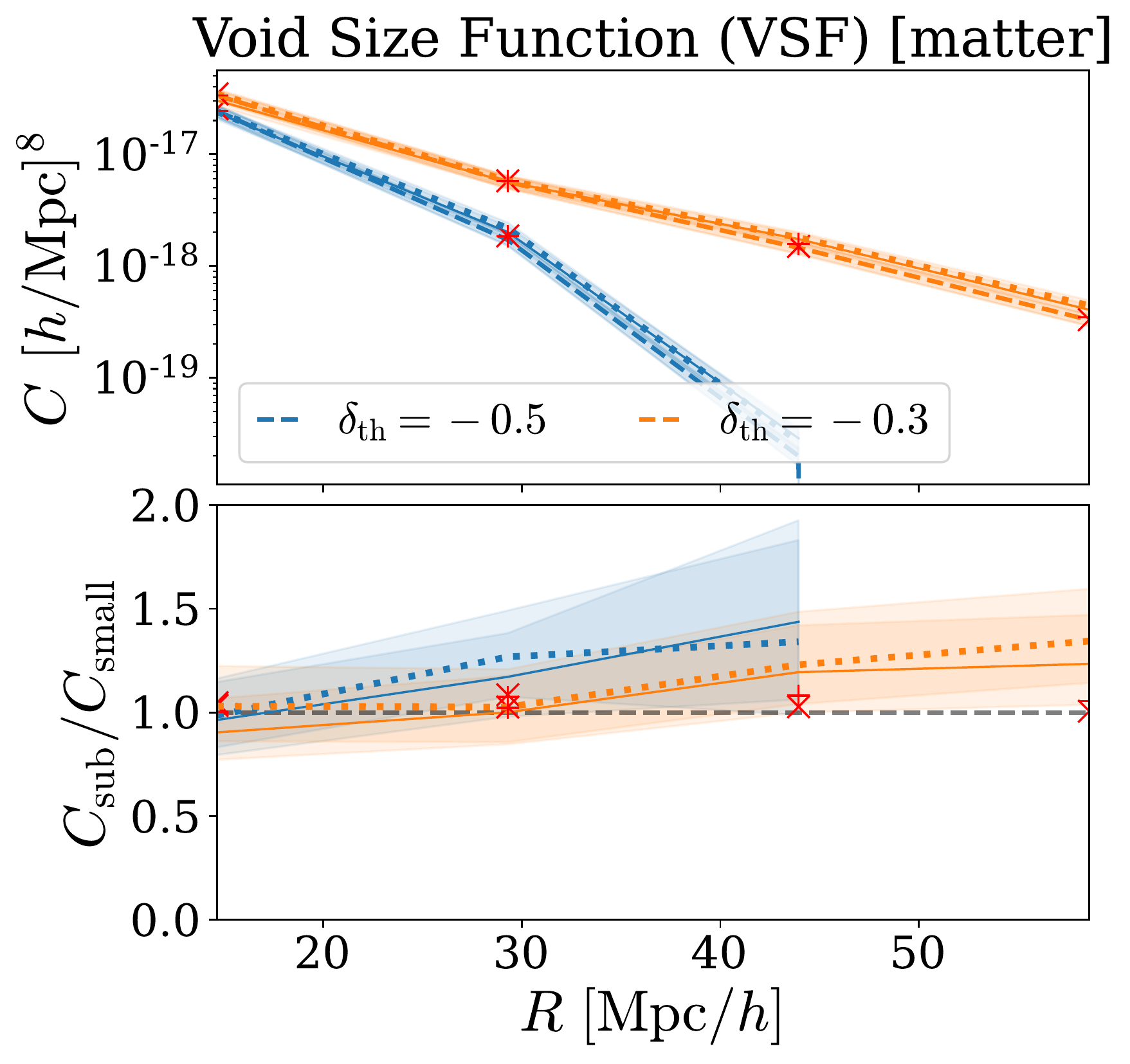}
    \vspace{.5ex}
    \includegraphics[width=0.32\textwidth]{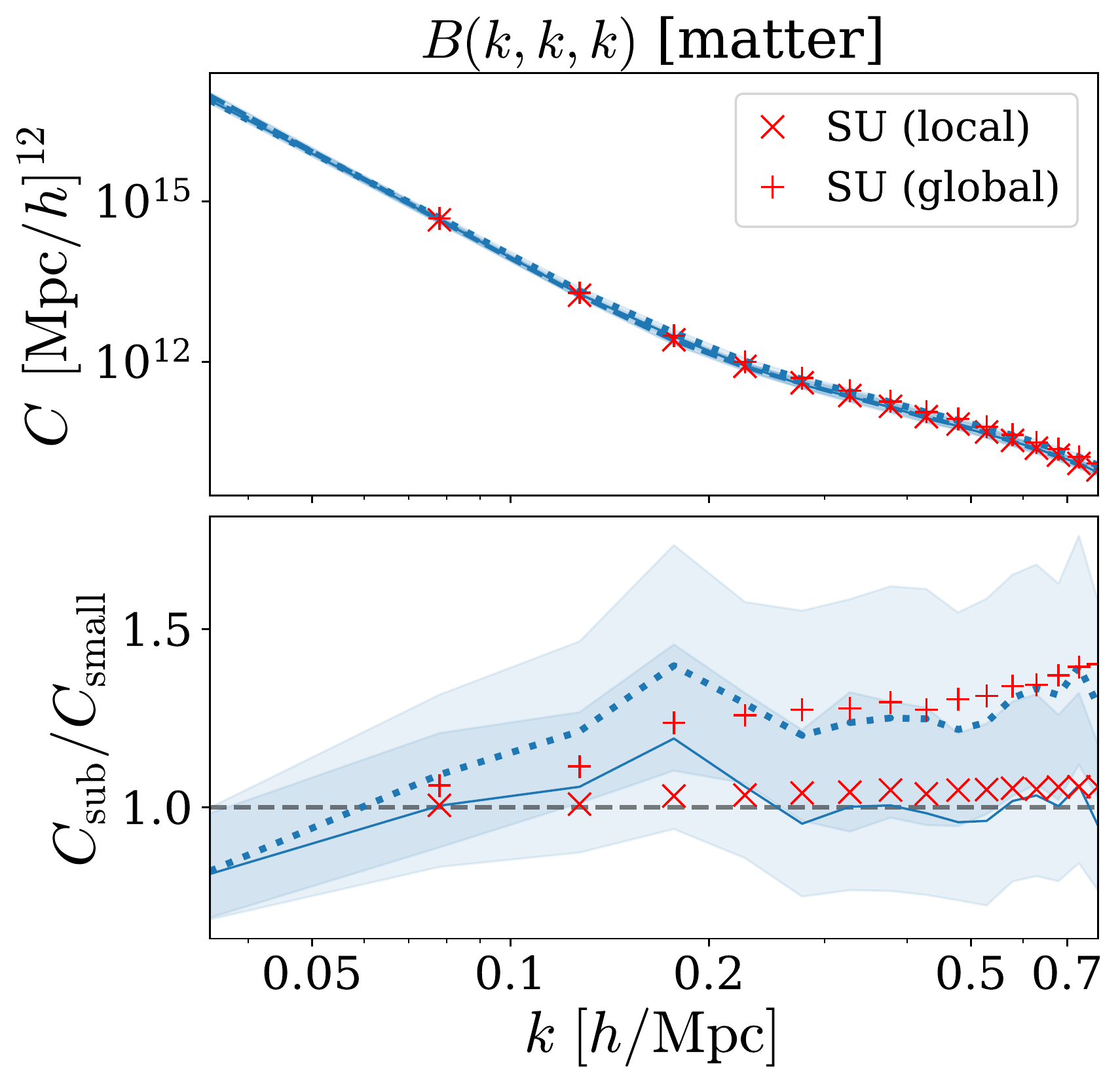}
    \includegraphics[width=0.32\textwidth]{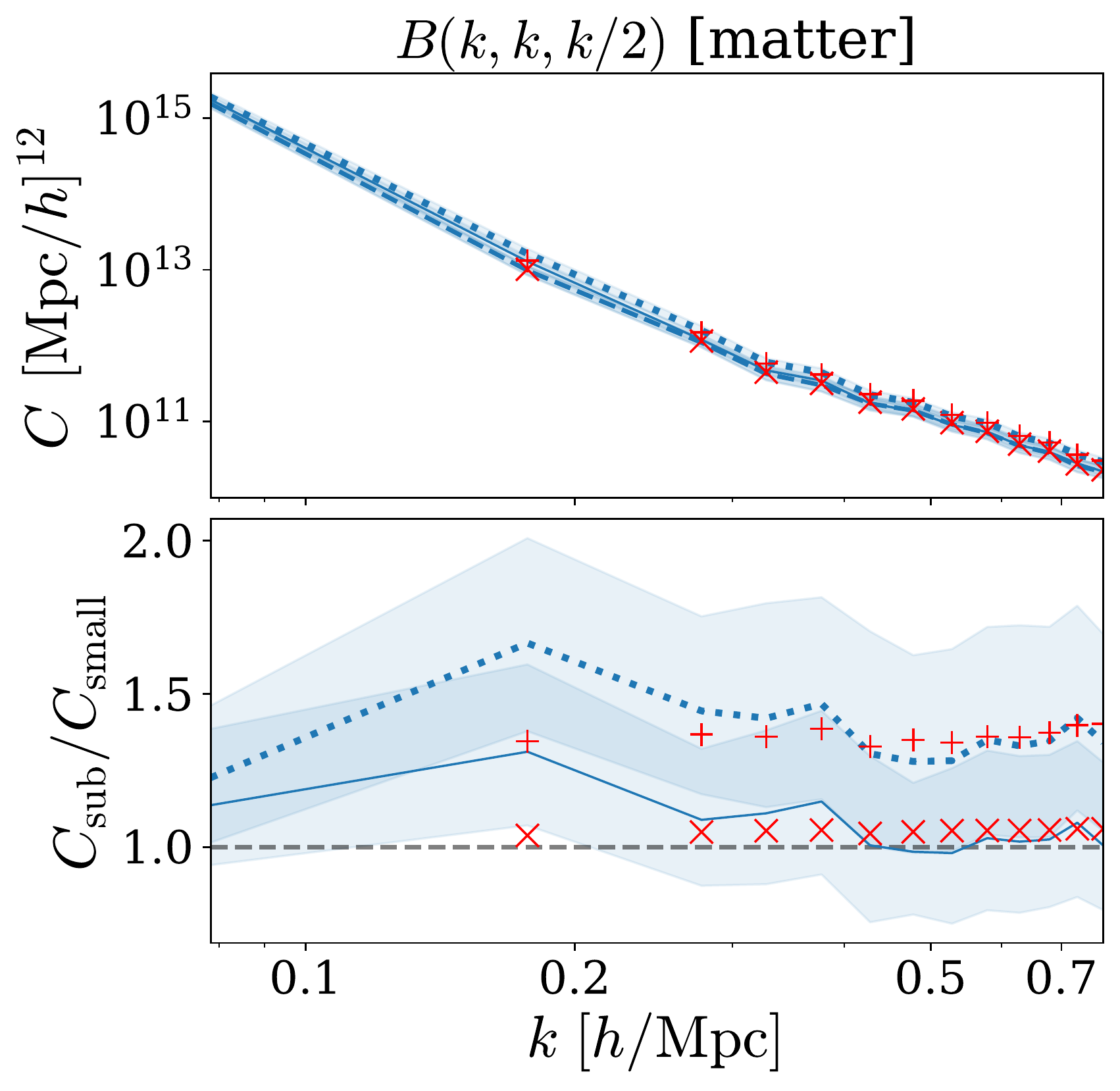}
    \includegraphics[width=0.32\textwidth]{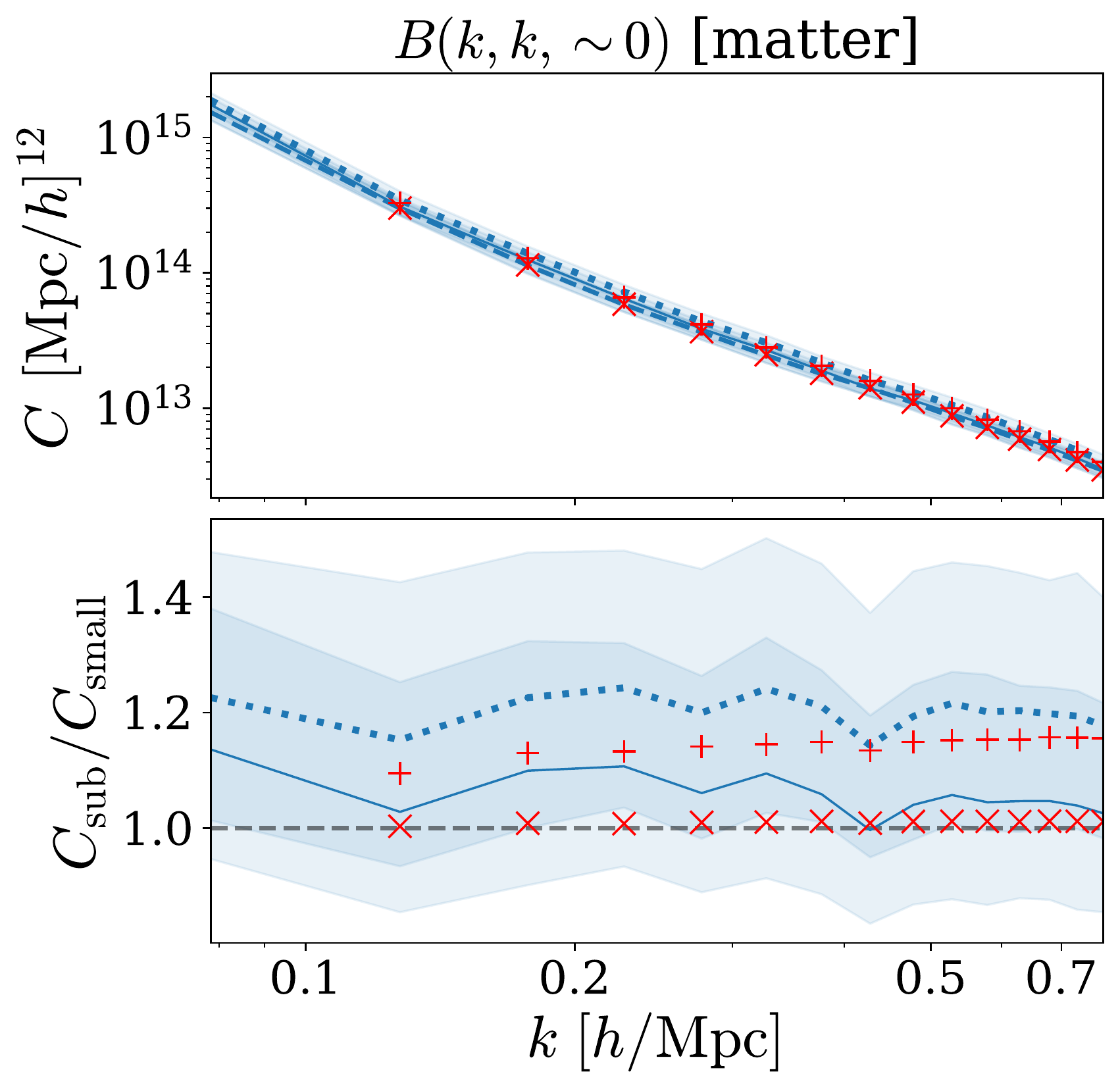}
  \end{subfigure}
  \caption{Covariances for the power spectrum (top left), halo mass function (top middle), void size function (top right), and three bispectrum configurations (bottom) in the matter field. Each statistic contains two panels, the top of which shows the diagonal term of the covariance computed in the small box (dashed), and in the sub-box using local (solid) and global (dotted) mean. The lower panel shows the ratio between the sub- and small boxes, where the dashed horizontal grey line indicates no SSC. For the void size function, we show results for two density thresholds used for void searching, $\delta_{\rm th}=-0.5$ (blue) and $\delta_{\rm th}=-0.3$ (orange). We also show separate universe results for SSC computed using local and global mean densities, marked in ``x'' and ``+'', respectively.  Shaded regions correspond to $95\%$ confidence intervals.}
\label{fig:var_cdm} 
\end{figure*}

Fig.~\ref{fig:var_cdm} shows the results of the SSC contribution for the power spectrum, void size function, halo mass function, and three bispectrum configurations. Each statistic contains two panels, the upper of which shows the diagonal term of the covariance computed with and without SSC, using sub- and small boxes, respectively. For sub-boxes, we show results using both the local mean  and the global mean density. The lower panel shows the ratio between the sub and small box, which is equal to $C_{\rm sub}/C_{\rm small} = 1+ C_{\rm SSC}/C_{\rm small}$ (using Eqn.~\ref{eq:C_SSC}). Shown also is the result from the SU approach (marked by the red crosses and pluses for the local and global mean cases respectively), which agrees reasonably well overall with the SSC estimated from the sub-box approach.

For the case of our power spectrum and bispectrum results, we note also that while the SSC does not depend to first order on the size of the wavenumber bins, other contributions to the covariance typically do, which can have an impact on the exact values of $C_{\rm sub}/C_{\rm small}$ (note this does not mean there is a dependence of the signal-to-noise on the bin size). This does not have an impact on the main takeaway points of our results, but it is useful to keep in mind especially when comparing quantitatively to results obtained previously in the literature.

\subsubsection{Power spectrum}
\label{sec:ps}
The SSC of the matter power spectrum can be seen to be a just under a 100\% effect at $k=0.7\,h/{\rm Mpc}$ in the local mean case. However, the SSC is much larger when using the global mean, with the ratio $C_{\rm sub}/C_{\rm small}$ increasing with $k$ to a factor of $\sim 4$ at $k=0.7\,h/{\rm Mpc}$. 
This can be explained as follows: the local mean density is modified with respect to the global mean by the background density $\delta_b$, as $\bar{\rho}_{\rm local} = \bar{\rho}_{\rm global} (1+\delta_b)$. Thus the power spectrum with respect to the local mean is given by $P_{\rm local}(k) = (1+\delta_b)^{-2}P_{\rm global}(k)$, where $P_{\rm global}(k)$ is the power spectrum with respect to the global mean.
The local and global responses are then related as,
\begin{equation}
    \frac{d \ln P_{\rm local}(k)}{d\delta_b} \approx \frac{d \ln P_{\rm global}(k)}{d\delta_b}-2,
    \label{eqn:Pm_response}
\end{equation}
where we use the fact that $\delta_b  = 0.03 \ll 1$.
It can be shown with perturbation theory that the global response is close to 2 for the scales considered in this paper \cite{dePutter_2012,Takada_2013,Li_2014}, hence the local response is much suppressed in comparison to the global. Recall, the SSC of the power spectrum referenced to the global density is what is relevant to weak lensing analysis, and this strong response is ultimately the reason why SSC is the most important piece of the off-diagonal covariance in cosmic shear 2-point function studies \cite{Barreira_2018, Barreira:2018jgd}.



\subsubsection{Halo Mass Function}

We find the SSC has very little contribution to the counts of massive halos $\gtrsim 10^{15} M_\odot / h$, while it increases towards less massive halos, with the ratio $C_{\rm sub}/C_{\rm small}$ becoming roughly a factor of 3 for masses of $\sim 10^{14} M_\odot/h$. 
The fluctuation in the number density of halos $\delta n = \delta n(M)$ 
in a mass bin $M$ 
is a biased tracer of the underlying matter field $\delta_m$
\begin{align}
\frac{\delta n}{\bar{n}} =  b \delta_m,
\end{align}
where $b=b(M)$ 
is the halo bias in the mass bin. 
The diagonal term of the sub-box HMF covariance divided by the shot noise $C_{\rm small} = \bar{n}/V$ is thus~\cite{Hu_2003},
\begin{align}
    C_{\rm sub} / C_{\rm small} = 1+ \sigma^2_b b^2 \bar{N}, 
    \label{eqn:Rcov_abun}
\end{align}
where $\bar{N}=\bar{n}V$ is the number (or abundance) of halos.
The second term on the right is the SSC contribution. While massive halos tend to be more biased (by a factor of few compared to low mass halos), their abundance is exponentially suppressed. Thus, the most massive halos are in the shot-noise dominated regime with little contribution from SSC.


\subsubsection{Void size function}
\label{sec:voids}

Recall, we consider spherical voids with density thresholds $\delta_{\rm th} = -0.5$ and $-0.3$, whose results are shown on the top right panel of Fig.~\ref{fig:var_cdm}. The SSC contribution for voids is generally small for all void radii shown.
Following from the discussion above for the HMF, this is as expected since voids are approximately 100 times less abundant than the halos, and their bias values remain of order $0-10$ \cite{Jamieson_2019, Chan_2020}. Thus, the covariance is strongly dominated by shot noise and the SSC effect is negligible. Note further that for certain void radii, and contrary to the case of halos, the void bias can be zero, in which case even an infinite abundance would have no SSC. 

It can, however, be seen that the error bars from the small versus sub-box analysis allows for an $\mathcal{O}(10-100\%)$ SSC effect, with a slight increase with $R$, while the SU approach suggests a negligible SSC for all $R$. This is likely a numerical artifact due to the void finder assuming periodic boundary conditions: the covariance of the sub-boxes, which are not periodic, could be spuriously large, especially for large voids that extend beyond the boundary of the box. On the other hand, the SU results are based on the difference between two periodic boxes, and are thus free of such boundary effects. We leave correcting this boundary effect for future work.

\subsubsection{Bispectrum}

All three configurations of the bispectrum that we consider have a much smaller SSC than the power spectrum, typically a factor 1--2 effect for $k \lesssim 0.7 h/{\rm Mpc}$. Similarly to the power spectrum, the global case has a larger effect than local. This agrees within error bars with~\cite{Chan_2018}, which considered the equilateral and isosceles cases using an analytical response approach and simulations. Also \cite{Coulton_2022} found the SSC effect to be small for the matter bispectrum. Our results are also in good agreement with~\cite{Barreira_2019_squeze} which analytically derived that the SSC effect is small for the squeezed matter bispectrum, as a few multiplicative factors get suppressed by the soft (low $k$) modes. 
More generally speaking, one can intuit that the matter bispectrum has a smaller SSC than the power spectrum as follows. While the power spectrum only has a Gaussian piece and a connected 4-pt function non-Gaussian piece (which includes the SSC), the bispectrum consists of additional non-Gaussian disconnected pieces (proportional to the square of the bispectrum and the power spectrum times the trispectrum) which increase the covariance overall. Thus, the connected 6-pt function (which is where the bispectrum's SSC enters) is small in comparison to the lower-order contributions to the bispectrum covariance.

\subsection{Halo field statistics}
\label{subsec:halos}

\begin{figure*}[t!]
\centering
  \begin{subfigure}[t]{\textwidth}
    \includegraphics[width=0.32\textwidth]{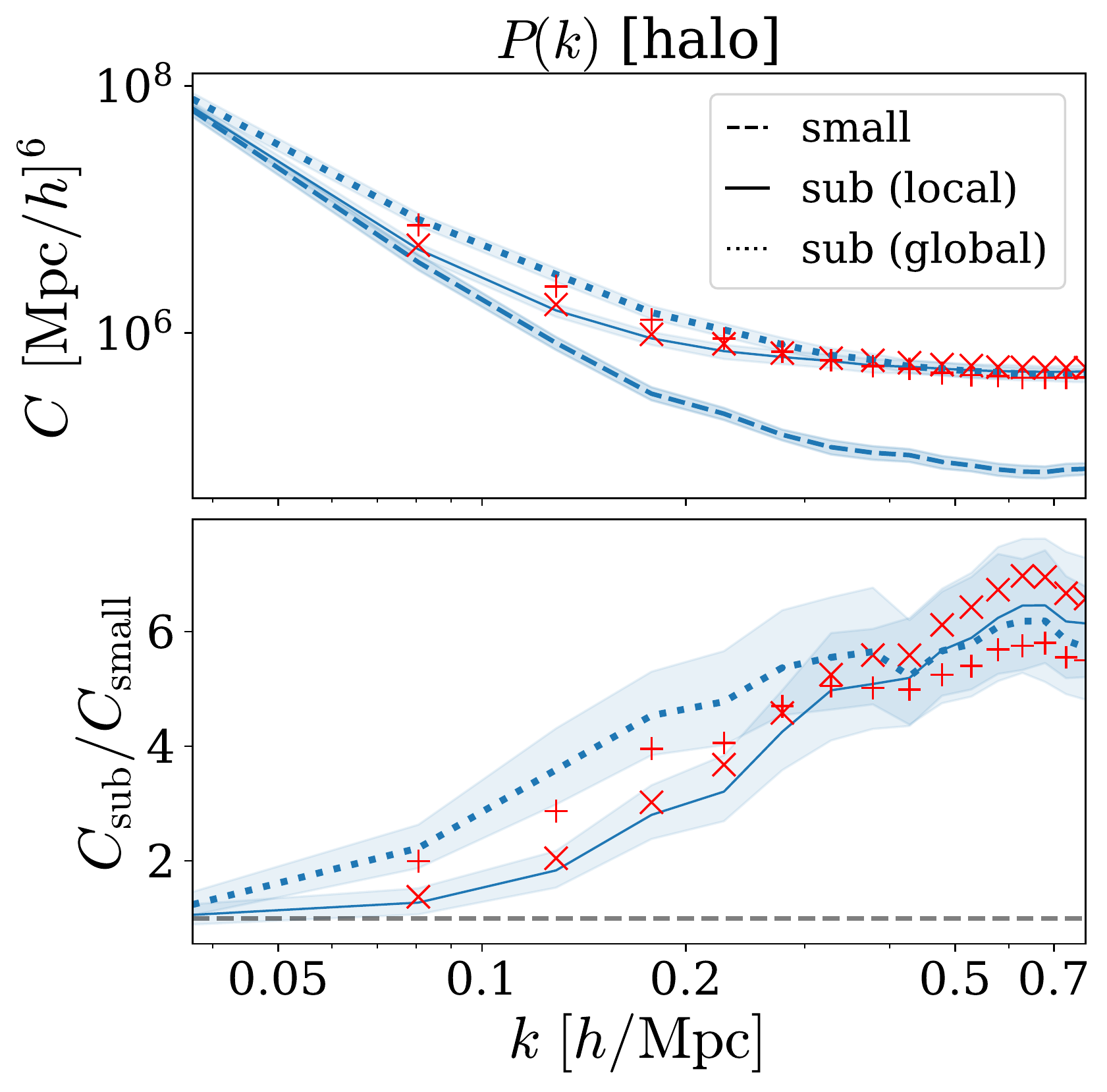}
    \includegraphics[width=0.32\textwidth]{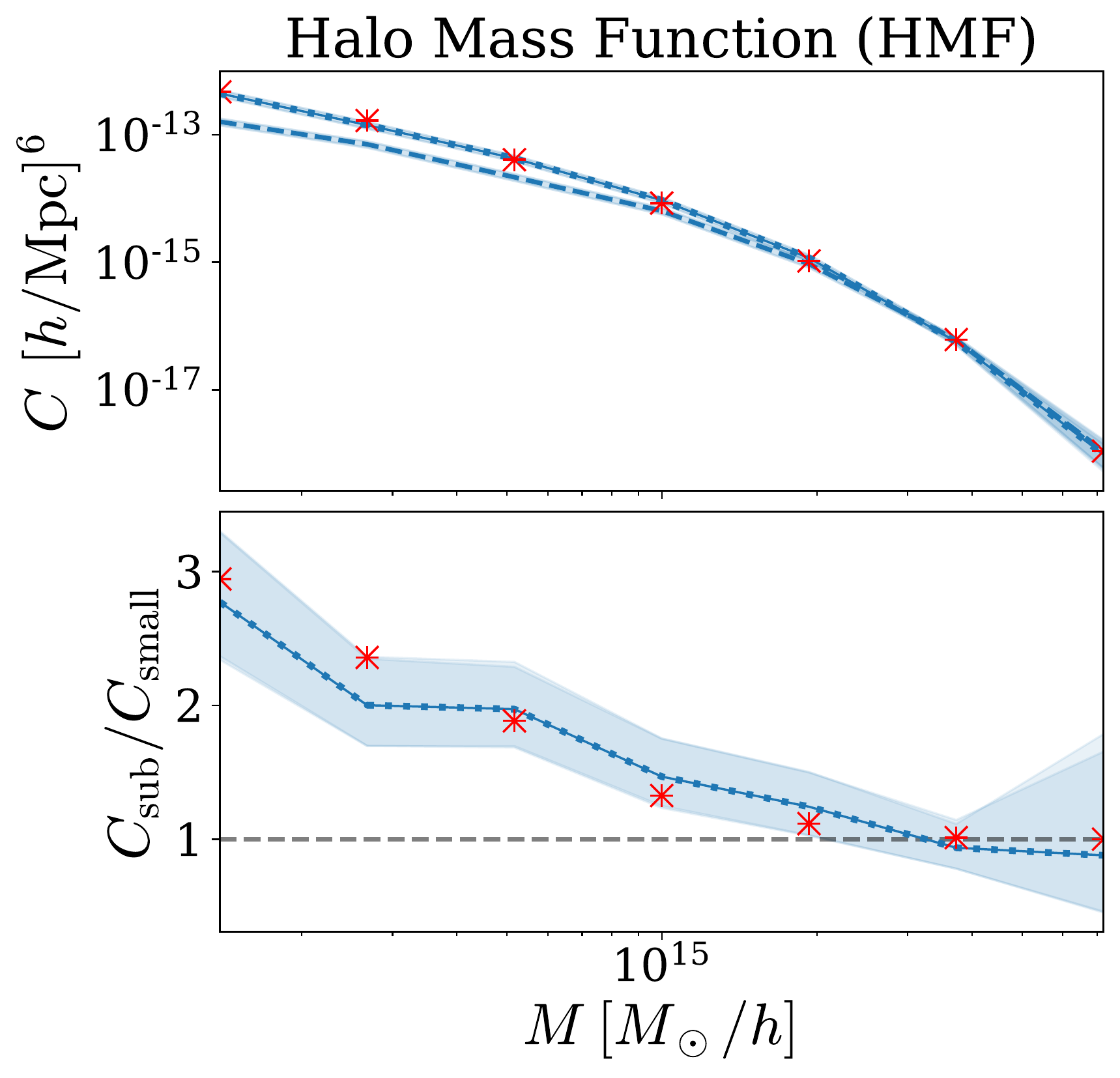}
    \includegraphics[width=0.32\textwidth]{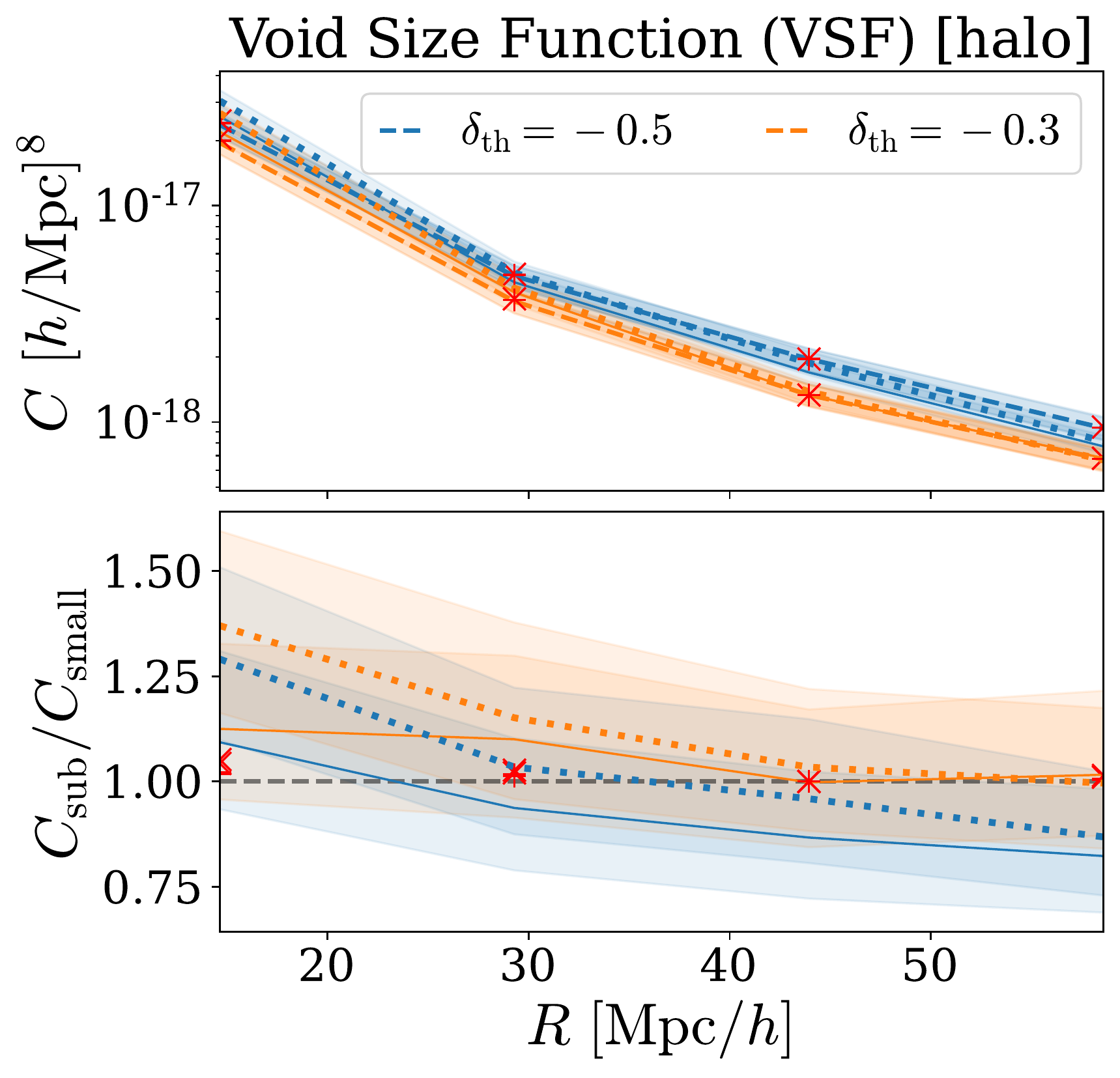}
    \vspace{.5ex}
    \includegraphics[width=0.32\textwidth]{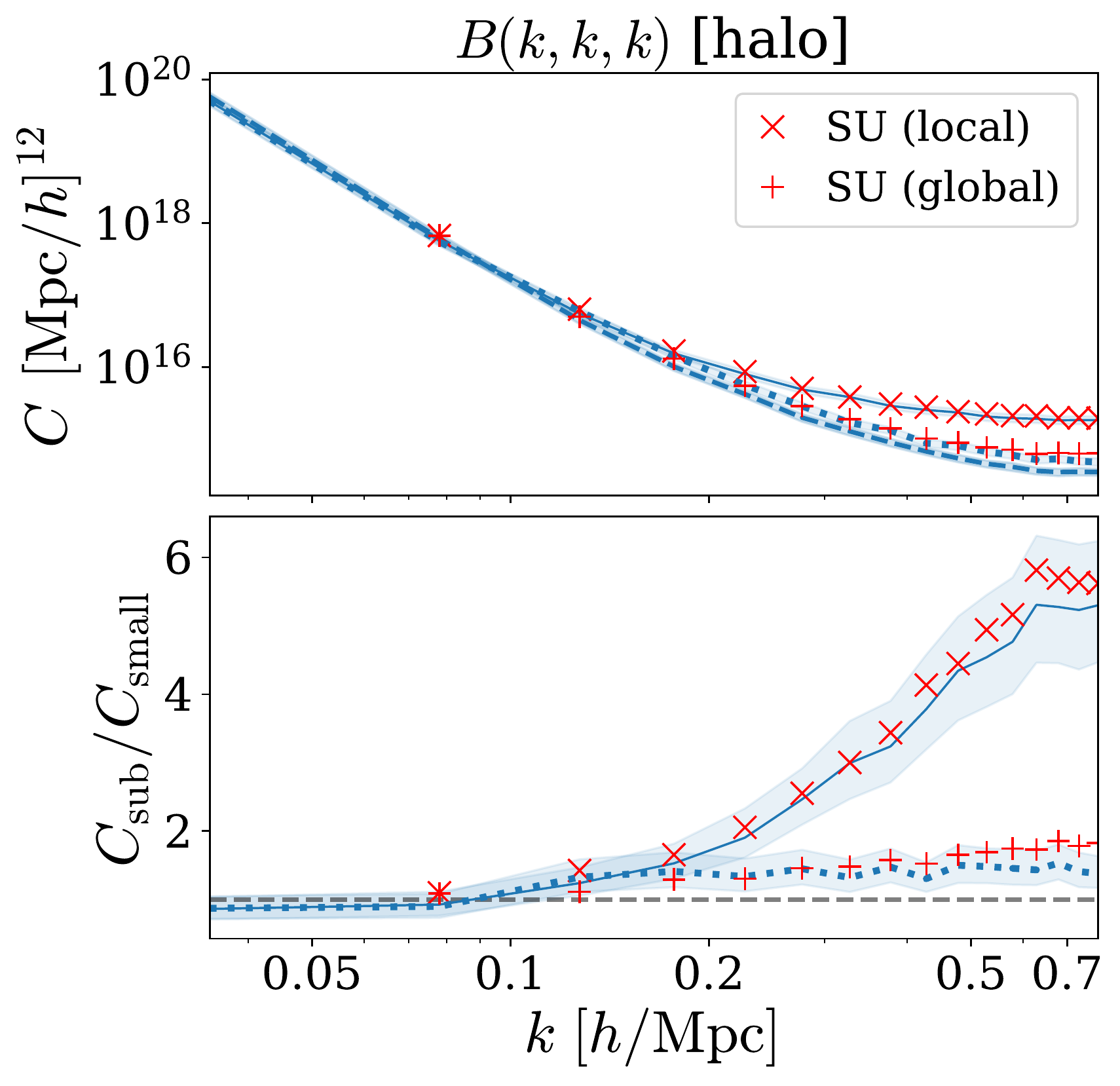}
    \includegraphics[width=0.32\textwidth]{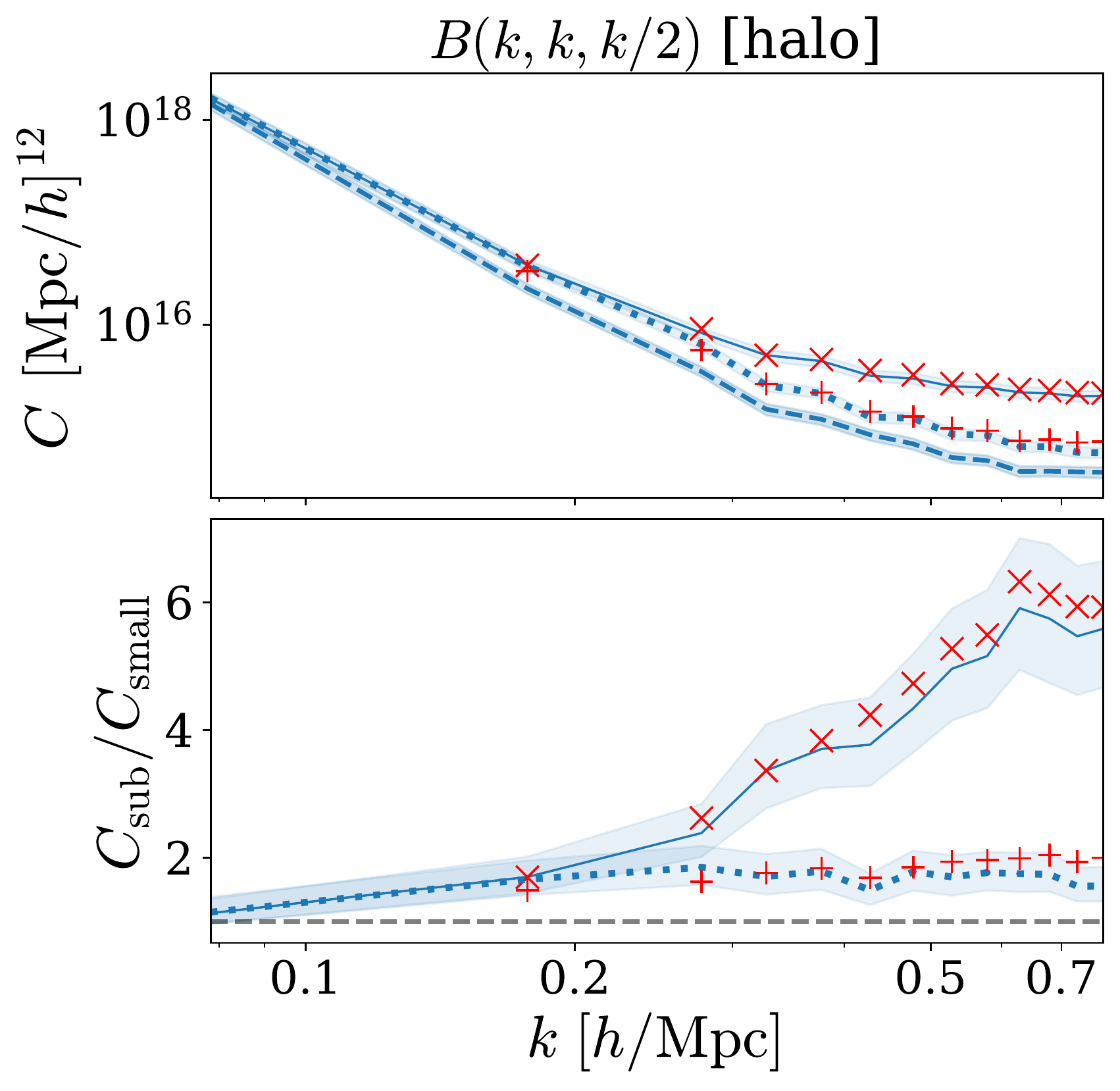}
    \includegraphics[width=0.32\textwidth]{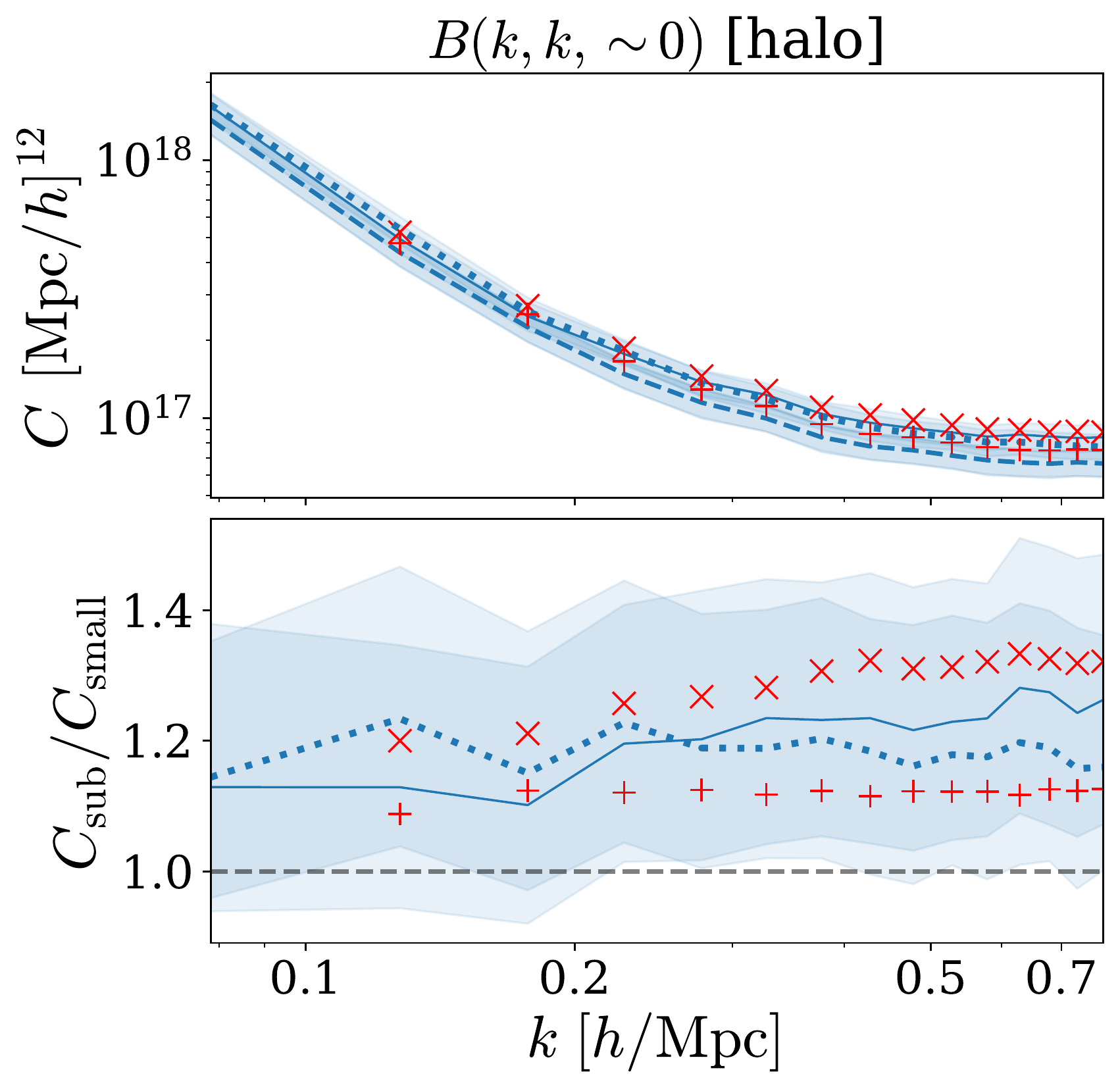}
  \end{subfigure}
  \caption{Same as Fig.~\ref{fig:var_cdm} but for the halo field, with a minimum mass cut $M_{\rm min} = 10^{14} M_\odot / h$. The halo mass function plot (top middle) is duplicated here for completeness.
  }
\label{fig:var_fof_N} 
\end{figure*}

\begin{figure*}[t]
\centering
    \includegraphics[width=\linewidth]{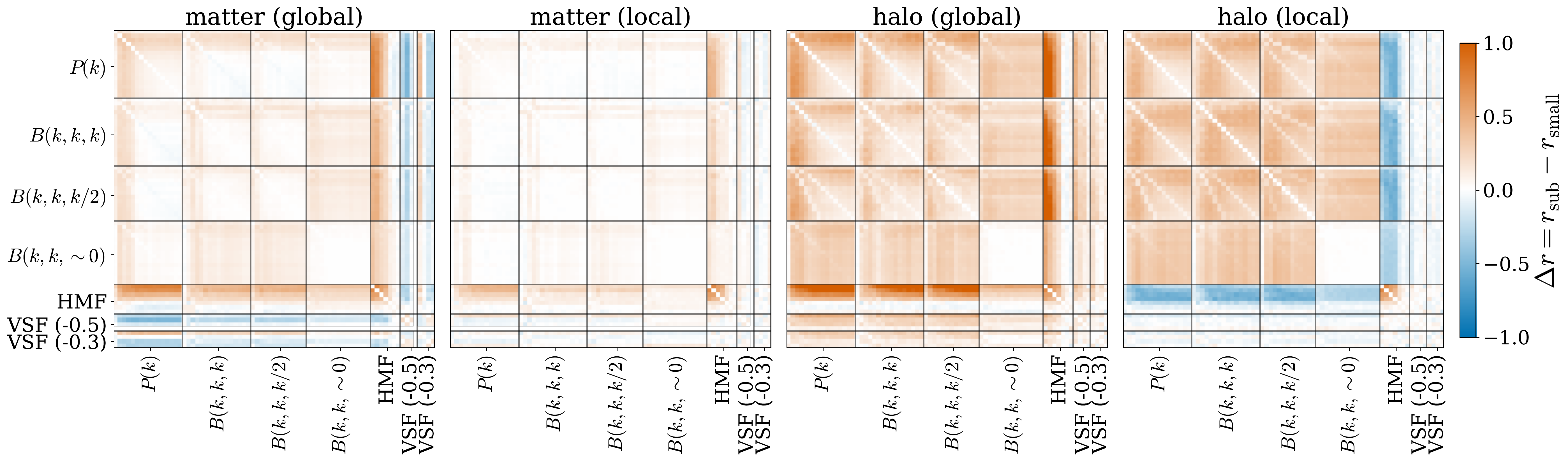} 
  \caption{Difference in correlation coefficient between the sub and small box $\Delta r \equiv r_{\rm sub} - r_{\rm small}$. From left to right: matter global, matter local, halo global, halo local.}
\label{fig:D_corr} 
\end{figure*}

Fig.~\ref{fig:var_fof_N} shows the results for statistics computed from the halo field, which is a biased tracer of the matter field. Here, we apply a minimum mass cut of $M_{\rm min} = 10^{14} M_\odot / h$.  
The halo field SSC shows qualitative similarities to that of the matter field, namely larger SSC on smaller scales for the power spectrum and the bispectrum, and a small SSC for voids.

One noteworthy difference concerns the relative size of the local and global cases for the power spectrum, which is now comparable.
This is because the local and global responses for halos are related as 
\begin{equation}
    \frac{d \ln P^h_{\rm local}(k)}{d\delta_b} \simeq \frac{d \ln P^h_{\rm global}(k)}{d\delta_b}-2b.
\end{equation}
Unlike in the case of the local matter response (Eqn.~\ref{eqn:Pm_response}), where the global response nearly cancels with the $-2$ term,
the global halo response is corrected by $-2b_1$. In our work the bias is  $b_1\approx2.5$, which leads to a negative local halo response (see Fig.~12 of \cite{Baldauf_2016} and derivation therein), so much so that the local effect is now comparable to the global effect after taking the square of the response in Eqn.~\ref{eqn:C_SU}. The exact value of the response is mass and redshift dependent, as halo bias increases when considering more massive halos and/or halos at higher redshift.
The SSC of the halo bispectrum will be similarly affected by the bias and mass cuts, the exact manner in which requires future theoretical study. For the setup considered in this work, we find that the global halo bispectrum SSC remains small for all configurations, while the local SSC becomes large for the equilateral and isosceles configurations.


\subsection{Cross covariance}
\label{subsec:cross-cov}

In this section, we study the SSC contribution to the cross covariance of the statistics. Higher-order statistics typically have large off-diagonal terms in the covariance compared to the power spectrum and are usually studied jointly with other statistics. Therefore, it is important to study not only the variances of individual statistics, but also their cross covariance. 
To focus on the off-diagonal terms, we normalize the covariance using the diagonal term to obtain the correlation matrix
\begin{equation}
    r_{ij} \equiv C_{ij}/\sqrt{C_{ii}C_{jj}},
\end{equation}
where $C$ is the covariance matrix with indices run through the bins of all the statistics studied here. 
Fig.~\ref{fig:D_corr} shows the difference in correlation coefficient between the sub- and small box, $\Delta r \equiv r_{\rm sub} - r_{\rm small}$, for the both the matter and halo fields, using local and global mean densities.

For the power spectrum and the bispectrum, the SSC contribution to the cross covariance is positive in all cases. The amplitude is the smallest in the local matter case, due to the local mean response cancellation discussed in Sec.~\ref{sec:ps}. The local and global halo cases see comparable contributions, also similarly to what was observed for the diagonal term in Sec.~\ref{subsec:halos}. 


For the HMF, we see a large contribution of SSC to the cross covariance with other statistics. The effect is positive in all cases, except for the halo local-mean case.

For the VSF, we observe a relatively small contribution of SSC to the cross covariance with other statistics, except for the global-mean matter field. This is consistent with the observation of almost negligible SSC in VSF variances in Sec.~\ref{sec:voids}, where we discussed that the SSC is low due to the low abundance of voids.

\section{Conclusions}
\label{sec:conclusions}

We study the effect of SSC on the power spectrum, bispectrum, halo mass function, and void size function, as well as on the cross covariance between them. We consider both the total matter and halo fields. We compare the covariance that includes the SSC (computed using 625 Mpc/$h$ \textit{sub}-boxes of a 5 Gpc/$h$ simulation), to the covariance without (computed using 625 Mpc/$h$ \textit{small} periodic boxes). We now summarize our main results together with additional discussion: 
\begin{itemize}
    \item We see an increasing impact of SSC on smaller scales for the matter and halo power spectrum, reaching a factor of a few compared to the covariance that ignores SSC from $k\approx 0.2 h/$Mpc and beyond. The exception is the case of the 
    power spectrum referenced to the local-mean density, for which the effect is less than $100\%$. This is in agreement with previous studies \cite{Hamilton_2006, dePutter_2012, Takada_2013, Li_2014, Takahashi_2009}.
    \item For the halo mass function, the SSC has little effect on massive halos above $10^{15} M_\odot / h$, as they are dominated by shot noise. However, the effect of SSC increases to a factor of 2--3 for lower mass halos, consistently with previous studies~\cite{Hu_2003, Schaan_2014, Philcox_2020_ehm}.
    \item For the void size function, we found a relatively small SSC effect, due to the low number density of voids compared to that of halos for a given survey volume, and also the low bias. This is an attractive feature of voids, making their covariance simpler to approximate without considering the SSC. 
    In this work we considered spherical voids, however one might consider different void finding algorithms, such as \texttt{VIDE}~\cite{VIDE}. However, as long as the void abundance and bias are comparable to those we considered, our general conclusions should hold for other void definitions as well.
    \item While the {\it matter} bispectrum receives less SSC contribution, an approximately 50\% effect, in good agreement with previous studies~\cite{Chan_2018,Barreira_2019_squeze}, the {\it halo} bispectra shows  dependence on the bispectrum configuration and the choice of local or global mean density. Concretely, the level of SSC remains relatively low ($\approx$10\%--100\% level) for all three halo bispectrum configurations (equilateral, isosceles, and squeezed) when using the global-mean density and for squeezed bispectrum using the local mean. However, the SSC contribution is a many hundred percent effect for the equilateral and isosceles halo bispectra referenced to the local mean.
    \item For the cross covariances, we see non-negligible contribution of SSC, in particular for the halo field statistics. We also observe a negative effect of SSC (or reducing the off-diagonal terms) for HMF $\times$ other statistics in the halo-local mean case, and VSF $\times$ other statistics in some radius bins. This indicates the importance of including the effect of SSC in future joint-statistic analysis.
    
\end{itemize}

In summary, our work shows that future cosmological analyses with the power spectrum and higher-order statistics, as well as their joint analysis, should need to carefully consider the effect of SSC. Analyses where many higher-order statistics are combined can have large data vector sizes, which puts
pressure on simulation-based methods for the covariance because of the need to have a sufficiently converged covariance matrix that is stable under inversion (which is what is needed in parameter inference analyses). This therefore strongly motivates more simulation-based works like ours here towards a robust understanding of the super-sample covariance and cross covariance of higher-order statistics. 

The level of impact of SSC depends on the box/survey size, halo sample, and redshift -- it would be fruitful future work to investigate these dependencies in detail. 
In particular, we considered a box size of $625 \,{\rm Mpc}/h$, a value in the ballpark of simulation volumes currently used for covariance estimation (see e.g.~Table 1 of \cite{Kacprzak:2022pww}). To translate our results to different survey volumes, the scaling of the super-sample covariance with window volume can be determined from Eqn.~\ref{eqn:C_SU}. The responses to $\delta_b$ are volume independent, thus the $\sigma_b^2$ term is solely responsible for volume dependence. Translating between different window volumes thus simply requires reevaluating $\sigma_b$. To gain some intuition, it can be seen that if the power spectrum were taken out of the integral in Eqn.~\ref{eqn:sigma_b}, the scaling would be exactly $C_{\rm SSC}\sim\sigma_b^2\sim1/V_W$, just like for other covariance contributions. However, performing the integral of $P_{\rm lin}$ convolved with the window function causes some corrections to this $1/V_W$ scaling.

It would also be interesting to assess the impact of the SSC at the level of final parameter posteriors in simulated likelihood inference analyses for ongoing and future surveys. Furthermore, it would be fruitful to use our simulations to quantify the SSC of other higher-order statistics. 
To enable such future works, our simulations have been made publicly available at \href{https://github.com/HalfDomeSims/ssc}{\faGithub}\footnote{\url{https://github.com/HalfDomeSims/ssc}}.



\begin{acknowledgments}
We thank Masahiro Takada, Eiichiro Komatsu, Yue Nan, Uroš Seljak, James Sullivan, and Francisco Villaescusa-Navaro for insightful discussion. 
This work was supported by JSPS KAKENHI Grants 23K13095 and 23H00107 (to JL). AEB thanks Kavli IPMU for hosting him during the duration of this project. 
AB acknowledges support from the  Excellence  Cluster  ORIGINS  which  is  funded  by  the  Deutsche  Forschungsgemeinschaft  (DFG, German Research Foundation) under Germany’s Excellence Strategy - EXC-2094-390783311.
This research used resources of the National Energy Research Scientific Computing Center (NERSC), a U.S.~Department of Energy Office of Science User Facility located at Lawrence Berkeley National Laboratory, operated under Contract No.~DE-AC02-05CH11231 using NERSC award  HEP-ERCAP0023125.
This work was partly performed at the Aspen Center for Physics, which is supported by National Science Foundation grant PHY-1607611.
We use \texttt{FastPM}~\cite{Feng2016, Bayer_2021_fastpm} to simulate large-scale structure. 
We use \texttt{nbodykit}~\cite{Hand_2018} to compute overdensity fields and  power spectra. 
The bispectrum is computed using the \texttt{bskit} package~\cite{Foreman_2020}, which employs the FFT-based bispectrum estimators of~\cite{Scoccimarro_2000, Sefusatti_2016}.
We use Pylians3\footnote{\url{https://pylians3.readthedocs.io/}} \cite{Pylians_2018} to find voids and compute the void size function.

\end{acknowledgments}

\bibliographystyle{physrev}
\bibliography{references}

\end{document}